\documentclass[%
reprint, 
bibnotes,
amsmath,amssymb,
aps,
]{revtex4-1}

\usepackage{graphicx}
\usepackage{dcolumn}
\usepackage{bm}
\usepackage{multirow}
\usepackage{color}
\usepackage{comment}
\definecolor{blue2}{rgb}{0,0.6,1}

\begin{document}
\title{Peta-Pascal Pressure Driven by Fast Isochoric Heating with Multi-Picosecond Intense Laser Pulse}

\author{Kazuki Matsuo$^{1}$*}
\author{Naoki Higashi$^1$}
\author{Natsumi Iwata$^1$}
\author{Shohei Sakata$^1$}
\author{Seungho Lee$^1$}
\author{Tomoyuki Johzaki$^2$}
\author{Hiroshi Sawada$^3$}
\author{Yuki Iwasa$^1$}
\author{King Fai Farley Law$^1$}
\author{Hiroki Morita$^1$}
\author{Yugo Ochiai$^1$}
\author{Sadaoki Kojima$^1$}
\author{Yuki Abe$^1$}
\author{Masayasu Hata$^1$}
\author{Takayoshi Sano$^1$}
\author{Hideo Nagatomo$^1$}
\author{Atsushi Sunahara$^4$}
\author{Alessio Morace$^1$}
\author{Akifumi Yogo$^1$}
\author{Mitsuo Nakai$^1$}
\author{Hitoshi Sakagami$^5$}
\author{Tetsuo Ozaki$^5$}
\author{Kohei Yamanoi$^1$}
\author{Takayoshi Norimatsu$^1$}
\author{Yoshiki Nakata$^1$}
\author{Shigeki Tokita$^1$}
\author{Junji Kawanaka$^1$}
\author{Hiroyuki Shiraga$^1$}
\author{Kunioki Mima$^1$, $^6$}
\author{Hiroshi Azechi$^1$}
\author{Ryosuke Kodama$^1$}
\author{Yasunobu Arikawa$^1$}
\author{Yasuhiko Sentoku$^1$}
\author{Shinsuke Fujioka$^1$*}
\affiliation{$^1$Institute of Laser Engineering, Osaka University, 2-6 Yamada-oka, Suita, Osaka 565-0871, Japan.}
\affiliation{$^2$Hiroshima University, 1-4-1 Kagamiyama, Higashi-Hiroshima 739-8527, Japan.}
\affiliation{$^3$Department of Physics, University of Nevada Reno, Reno, Nevada 89557, USA.}
\affiliation{$^4$Institute for Laser Technology, 1-8-4 Utsubo-honmachi, Nishi-ku Osaka, Osaka, 550-0004, Japan.}
\affiliation{$^5$National Institute for Fusion Science, National Institutes of Natural Sciences, 322-6 Oroshi, Toki, Gifu, 509-5292, Japan.}
\affiliation{$^6$The Graduate School for the Creation of New Photonics Industries, 1955-1, Kurematsu, Nishi-ku, Hamamatsu, Shizuoka 431-1202, Japan.}

\date{\today}
             
\begin{abstract}
Fast isochoric laser heating is a scheme to heat a matter with relativistic-intensity ($>$ 10$^{18}$ W/cm$^2$) laser pulse or X-ray free electron laser pulse.
The fast isochoric laser heating has been studied for creating efficiently ultra-high-energy-density (UHED) state. We demonstrate an fast isochoric heating of an imploded dense plasma using a multi-picosecond kJ-class petawatt laser with an assistance of externally applied kilo-tesla magnetic fields for guiding fast electrons to the dense plasma.The UHED state with 2.2 Peta-Pascal is achieved experimentally with 4.6 kJ of total laser energy that is one order of magnitude lower than the energy used in the conventional implosion scheme.
A two-dimensional particle-in-cell simulation reveals that diffusive heating from a laser-plasma interaction zone to the dense plasma plays an essential role to the efficient creation of the UHED state.
\end{abstract}

\maketitle


Power laser apparatus can inject a plenty of energy to a matter in a small volume within a short time duration. The matter becomes a high-energy-density state that is applicable to various scientific researches such as laboratory astrophysics \cite{Remington2006} and several kind of radiation sources: X-rays, charged particles, and neutrons \cite{Fujioka2009,Casey2017}. 
Fast isochoric heating of solid target to create such high-energy-density states have been demonstrated by short pulse laser \cite{Sawada2019} and X-ray free electron laser \cite{Vinko2015, Royle2017}.
The fast isochoric laser heating of imploded plasma is one of the scheme to create ultra-high-energy-density (UHED) state, which is equivalent to the state at the center of the sun, for the inertial confinement fusion (ICF) science.

In conventional implosion, kinetic energy of the imploded shell is converted to the internal energy of the compressed matter at the maximum compression.
A UHED state with 36 Peta-Pascal (PPa) was achieved on the National Ignition Facility with 1.8\,MJ of laser energy by the indirect X-ray driven implosion \cite{LePape2018}.
The OMEGA laser facility with 30\,kJ of laser energy produced 5.6 PPa of UHED by the direct laser-driven implosion \cite{Regan2016}.
The current central ignition scheme requires enormous laser energy to create UHED state. The significant growth of hydrodynamic instabilities during the compression causes the hot spark mixing with the cold dense fuel and prevents efficient creation of UHED state.

In the context of ICF, the fast isochoric heating also known as the fast ignition had been proposed as an alternative approach \cite{Tabak1994}. 
This approach separates compression and heating processes to avoid the mixing, using a more stable compression followed by an external energy injection whose time scale is much shorter than the implosion time scale. Hence, it potentially leads to higher gain by increasing the mass of the compressed fuel .

A cone-in-shell target is commonly used in the integrated fast-isochoric-heating experiments, since a dense plasma core, which is produced by the laser-driven implosion, is surrounded by a long-scale-length coronal plasma.
The cone preserves a vacuum channel in the coronal plasma for the heating laser pulse to access the compressed core.
Significant progresses have been made on the fast isochoric heating to increase the laser-to-core energy coupling \cite{Kodama2001, Theobald2014, Jarrott2016}.
Those improvements mainly focus on increasing laser-to-core energy coupling by drag heating which is the energy exchange through binary collisions between relativistic electrons (REs) and bulk electrons of the plasma. 

There are three major mechanisms of the fast isochoric heating.
The following equation describes temporal increment of electron temperature $T_e$ of a dense bulk plasma, whose electron number density is $n_e$ \cite{Kemp2006}.
\begin{equation}
\frac{3}{2}n_e\frac{\partial T_e}{\partial t} = \frac{3}{2}\frac{n_h T_h}{\tau_e(T_h)} + \frac{{j_h}^2}{\sigma(T_e)} + \frac{\partial}{\partial x}(\kappa(T_e)\frac{\partial T_e}{\partial x}). \label{heating}
\end{equation}
Here $\tau_e(T_h)$ is the collision time between bulk electrons and REs having $T_h$ of temperature in a fully ionized plasma. 
In the calculation of $\tau_e(T_h)$ the bulk electrons' temperature $T_e$ ($\ll T_h$) is neglected.
RE current is given as $j_h \simeq e n_h c$ with hot electron number density $n_h$, elementally charge $e$ and speed of light $c$.
Electric conductivity $\sigma$ of a plasma is $\sigma = n_c e^2 \tau / m_e$, where $n_c$, $\tau$, and $m_e$ are the critical density, 
collision time between bulk electons and ions, and the electron mass, respectively. 
$\kappa(T_e)$ is bulk electrons' thermal conductivity.

The first term of the right hand side is the drag heating. 
This mechanism contributes significantly to heating the over-solid density plasma when the energy of REs is optimized for them to be stopped in the plasma.
The REs' large divergence is a critical issue in the drag heating.
We had reported enhancement of the laser-to-matter energy coupling with the magnetized fast isochoric heating scheme \cite{Sakata2018}.
The maximum coupling efficiency via the drag heating reached 7.7 $\pm$ 1.2\% because of reduction of RE's divergence by the application of the external magnetic field.

The second term represents the resistive heating.
The RE current drives a return current $j_e$, whose flow direction is opposite to that of the RE current, to maintain current neutrality in the plasma, i.e. $j_e\simeq j_h$.
Since the return current is more collisional than the RE current, the return current heats the plasma ohmically.
The resistive heating is the dominant heating mechanism when the current density is high since $\propto j_h^2$. However, it becomes less effective when REs propagate with large divergence.

The third term is the diffusive heating, in which thermal electrons transport their energy diffusively from the laser heated hot region to the cold dense region.
The diffusive heating had not been considered in the previous integrated fast-isochoric heating experiments, where the dense plasma was heated with a sub-picosecond heating pulse.
We find that the diffusive heating becomes important for a multi-picosecond heating laser pulse.

In this study, we have created 2.2 PPa of UHED efficiently using a mulit-picosecond petawatt laser, LFEX, with the total 4.6 kJ energy (1.5kJ is for compression, 1.7kJ is for generating a magnetic field, and 1.4kJ is for heating laser) in the magnetized fast isochoric heating scheme.
The achievement of the UHED with the fast isochoric heating can not be explained only with the enhanced drag heating that we had report in Ref.\,\cite{Sakata2018}.
Two-dimensional particle-in-cell (2D-PIC) simulation reveals that the diffusive heating mechanism contributed significantly to the keV heating of an imploded plasma in a time scale of a few picoseconds \cite{Sawada2019}.

This achievement was realized with the following improvements.
The first improvement is the heating laser contrast.
A short pulse laser system emits inevitably weak pulses preceding the main pulse.
The intensity ratio of the main pulse to the preceding pulses is called the pulse contrast.
A low contrast laser system produces a long-scale-length plasma in the laser-plasma interaction zone before the main pulse arrival time.
The preformed plasma affects negatively on the dense plasma heating because of separation of the interaction zone away from the dense plasma.
The high contrast laser enables easy access to high density plasma.

The second is implementation of a solid ball target.
The solid ball compression does not generate shocks and rarefactions traveling ahead of the shock-compressed matter, therefore a cone tip is not required for preventing a hot plasma flowing into the cone \cite{Fujioka2015}.
The direct interaction of a dense matter with heating laser pulse is then realized with the open-tip cone, resulting enhancement of the diffusive heating from the interaction zone to the imploded dense plasma.

The third is a laser-driven capacitor coil target used for generating strong magnetic field \cite{Daido1986}. 
By applying external magnetic fields to the path of the REs and the thermalized electrons, which carry the heat, the divergence issue can be resolved \cite{Bailly-Grandvaux2018} so that the resistive heating near the interaction zone works efficiently to make a large temperature gradient for the diffusive heating.


\begin{table*}
	\caption{\label{table: shot_summary} Summary}
	\begin{tabular}{cccccccc}
		\hline
		\textbf{Case} & \textbf{Shot} & \textbf{Heating} & \textbf{Heating} & \textbf{Coupling} & \textbf{Electron} & \textbf{Mass} & \textbf{Pressure}\\
		 & \textbf{ID} & \textbf{Energy} & \textbf{Timing} & \textbf{} & \textbf{Temperature} & \textbf{Density} &\\
		 & & [J] & [ns] & [\%] & [keV] & [g/cm$^3$] & [PPa] \\\hline \hline
		A & 40558 & 1516 & 0.40 & 3.1 $\pm$ 0.5 & 2.1 $\pm$ 0.2 & 1.1 (+6.1 / -0.0) & 0.2 (+1.4 / -0.0)\\
		B & 40556 & 1016 & 0.61 & 4.3 $\pm$ 0.7 & 2.0 $\pm$ 0.1 & 11.3 (+5.9 / -2.2) & 2.2 (+1.3 / -0.5)\\
		C & 40547 & 1100 & 0.38 & 5.5 $\pm$ 0.9 & 1.9 $\pm$ 0.2 & 1.1 (+6.1 / -0.0) & 0.2 (+1.4 / -0.0)\\
		D & 40549 & 668 & 0.37 & 5.8 $\pm$ 0.9 & 1.5 $\pm$ 0.1 & 1.1 (+6.1 / -0.0) & 0.2 (+1.4 / -0.0)\\
		E & 40543 & 625 & 0.72 & 7.7 $\pm$ 1.2 & N/A & 11.3 (+5.9 / -2.2) & N/A \\\hline
	\end{tabular}
\end{table*}

Table \ref{table: shot_summary} summarizes the results in this experiment. 
The experiment had been conducted on GEKKO-LFEX laser facility at Institute of Laser Engineering, Osaka University. 
The laser conditions, laser-to-core coupling efficiency by the drag heating, the geometrical positions of the target, and the diagnostics were identical to those of our previous experiment \cite{Sakata2018}.
The experimental time origin ($t_{\rm exp}$ = 0 ns) is defined hereafter as the peak of the compression laser pulse.

Three of GEKKO-XII laser beams were used for driving a nickel-made capacitor-coil target.
Wavelength, pulse shape, pulse duration, and energy of the GEKKO-XII beams used for magnetic field generation were 1.053 $\mu$m, Gaussian, 1.3 ns full-width at half-maximum (FWHM), and 600 $\pm$ 20 J perbeam.
The strength of magnetic field generated with the capacitor-coil target had been measured on GEKKO-XII, LULI2000, Shengguang-II, and OMEGA-EP laser facilities \cite{Law2016, Fujioka2013, Zhu2015, Santos2015b, Gao2016}. 
600 - 700\,Tesla of magnetic fields were obtained by the current GEKKO-XII laser beams configuration.

Six of GEKKO-XII laser beams were used for compression of a solid ball target.
Wavelength, pulse shape, pulse duration, and energy of the GEKKO-XII beams used for implosion were 0.526 $\mu$m, Gaussian, 1.3 ns (FWHM), and 240 $\pm$ 15 J perbeam.
The target was made of a 200 $\mu$m-diameter Cu(II) oleate solid ball [Cu(C$_{17}$H$_{33}$COO)$_{2}$] \cite{Iwasa2017} coated with a 25 $\mu$m-thick polyvinyl alcohol (PVA) layer to prevent the Cu atoms from  directly laser irradiation.
The Cu(II) oleate solid ball contains 9.7$\%$ Cu atoms in weight. 
X-ray emissions from Cu atoms were used for characterizing the coupling efficiency and the electron temperature.
A open-tip gold cone was attached to the Cu(II) oleate solid ball.

\begin{figure}[b]
\begin{center}
 \includegraphics[width=75mm]{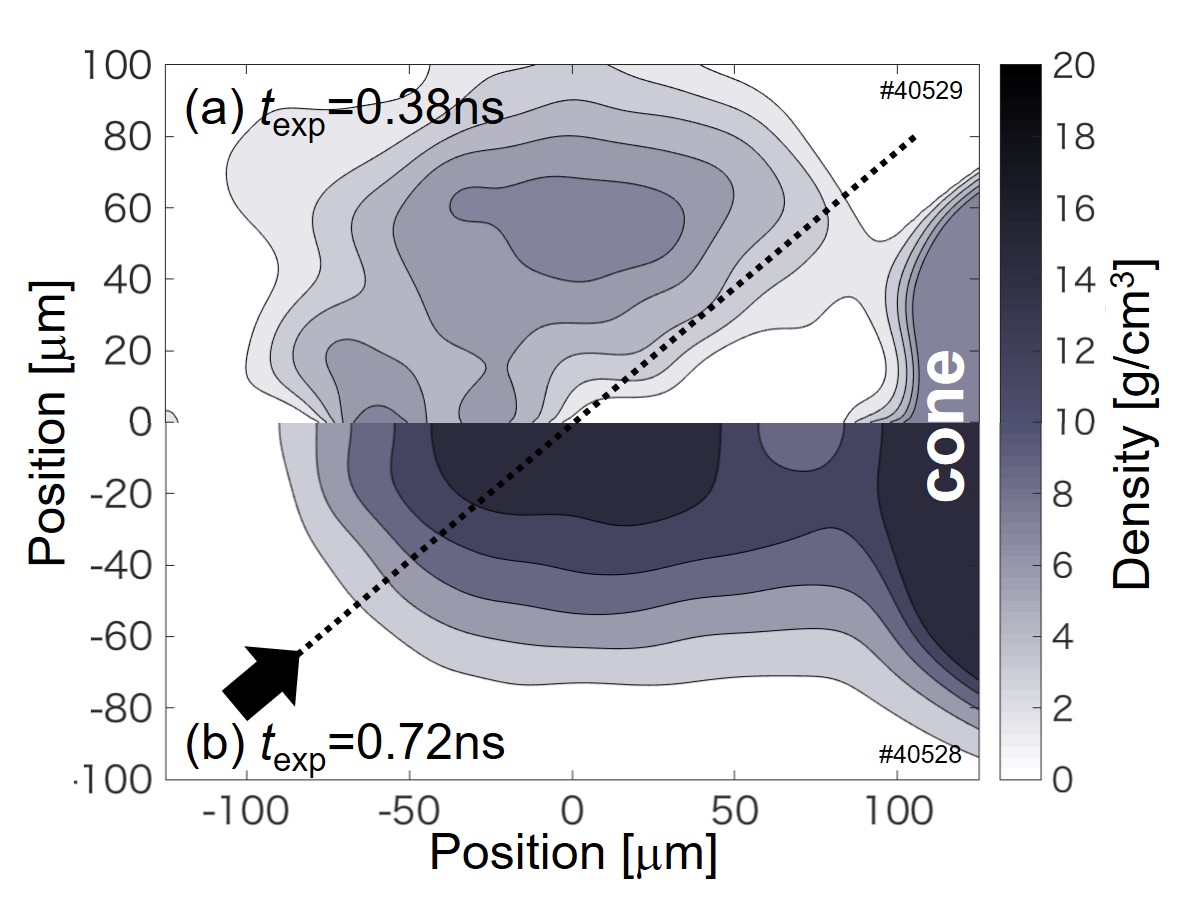}
\end{center}
 \caption{Two-dimensional density profiles of compressed Cu (II) oleate solid balls. (a) At $t_{\rm exp}$ = 0.38 ns which is before the maximum compression timing, the converging shock waves were still traveling to the center. (b) The core reached maximum compression timing at $t_{\rm exp}$ = 0.72 ns. The average mass density along the horizontal of the core is 11.3 $\rm g/cm^3$. Dotted lines show axis of the X-ray spectrometer.} \label{fig:density_map}
\end{figure}

Two-dimensional density profiles of compressed Cu(II) oleate solid balls were measured with flash X-ray backlight technique coupled with a spherically bent crystal X-ray imager \cite{Fujioka2010, Sawada2016}.
The magnification, spatial resolution, and spectral bandwidth were 20, 13 $\mu$m (FWHM), and 5 eV (FWHM), respectively.
The X-rays were passing through the compressed core and imaged by using a spherical bent crystal on an imaging plates. 
An X-ray shadowgraph is converted to an X-ray transmittance profile by interpolating the two-dimensional backlight X-ray intensity profile from the center of core to the outside region. 
The areal density of the pre-compressed core was calculated from the X-ray transmittance profile with a precomputed opacity of 100 eV Cu(II) oleate for 8.05-keV photon. 
A two-dimensional density profile of the core was then obtained after applying an inverse Abel transformation to the areal density profile, assuming rotational symmetry of the core along the cone axis as shown in Fig.\,\ref{fig:density_map}.
The converging shock waves were still traveling to the center of the solid ball at $t_{\rm exp}$ = 0.38 ns, therefore the area close to the cone maintained the initial density 1.1 $\rm g/cm^3$.
Maximum compression was reached at around $t_{\rm exp}$ = 0.72 ns. The average mass density of the core along the laser axis is 11.3 $\rm g/cm^3$.

Four LFEX beams were injected from the gold cone side to a compressed plasma at $t_{\rm exp}$ = 0.37 or 0.72 ns.
Wavelength, pulse shape, pulse duration, and energy of the LFEX beams were 1.053 $\mu$m, Gaussian, 1.8 $\pm$ 0.3 ps (FWHM), and varied from 620 to 1520 J.
The focal spot diameter was 50 $\mu$m (FWHM) containing 30$\%$ of the total energy, yielding an intensity of $1.3 \times 10^{19}$ $\rm W/cm^{2}$ at the maximum energy shot. 

\begin{figure}[t]
\begin{center}
 \includegraphics[width=75mm]{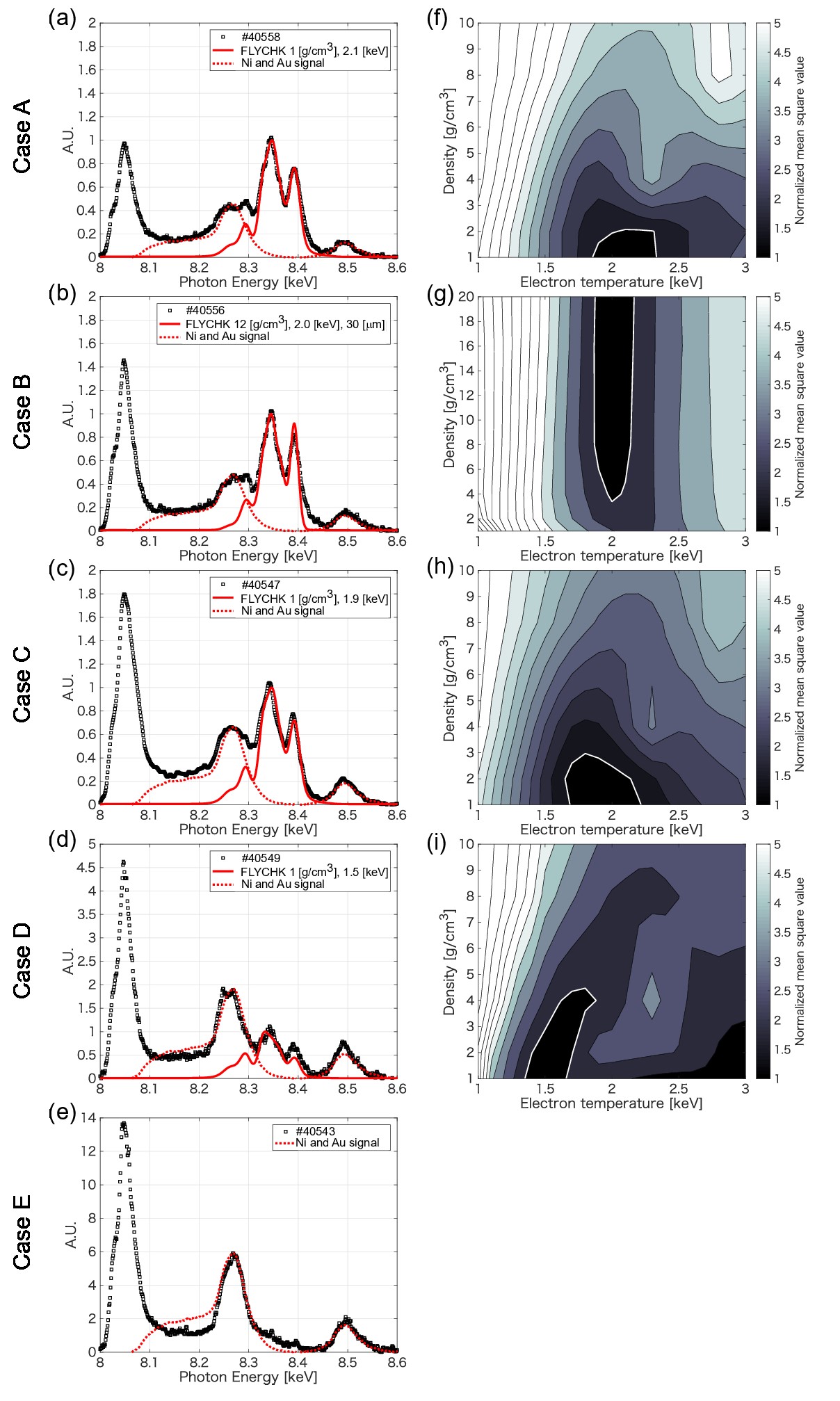}
\end{center}
\caption{X-ray spectra in the range from 8.0 to 8.6 keV and the least square values. 
(a), (b), (c), (d), (e)  Black square, red line, and red dotted line show the experimental spectra data, the computed spectra data by using FLYCHK code, 
and the Ni-$K_\beta$ and Au-$L$ lines subtracted from the measured spectra, (f), (g), (h), (i) The least square value is normalized by using the minimum least square value of each data. 
We can find the minimum least square value in the region surrounded by white line. 
The experimental data of Case A, C, and D were well reproduced with the density 1.0 $\rm g/cm^3$, and the temperature as 2.1 $\pm$ 0.2, 1.9 $\pm$ 0.2, 1.5 $\pm$ 0.1 keV, respectively. 
The experimental data of Case B were well reproduced with the density 12 $\rm g/cm^3$ and the temperature 2.0 $\pm$ 0.1 keV.} \label{fig:spectrum_result}
\end{figure}

The X-ray spectrometer was installed at 40 deg from the LFEX incident axis (dotted lines of Fig.\,\ref{fig:density_map}).
The spectrometer utilizes a planar highly oriented pyrolytic graphite (HOPG) as a dispersive element.
The spectral resolution of the spectrometer was 17.9 eV (FWHM). 
Figure \ref{fig:spectrum_result} (a,b,c,d,e) shows X-ray spectra in the range of 8.0 to 8.6 keV.
The peaks at 8.05, 8.35, and 8.39 keV are $\mathrm{Cu-}K_\alpha$, Li-like Cu satellite lines , and $\mathrm{Cu-}He_\alpha$, respectively.
The peak at 8.26 keV is Ni-$K_\beta$ that was emitted from a Ni-made capacitor-coil target.
The peak around 8.5 keV are Au-$L$ lines emitted from the Au cone.
The ratio of $\mathrm{Cu-}He_\alpha$ to Li-like satellite line reflects information on the electon temperature $T_e$, plasma density $\rho$ and thickness along the line of sight of the spectrometer $d$.

X-ray spectra from Cu were computed by using FLYCHK code \cite{Chung2005}. 
The Ni-$K_\beta$ and Au-$L$ lines were measured in separate shots, and these emissions were subtracted from the measured spectra after adjusting their peak intensity for the fitting. 
The thickness was determined so that the spectrum could be reproduced within the density from the X-ray backlight as Fig.\,\ref{fig:density_map}. 
The spectra were calculated with varying $\rho$ within the experimentally obtained density and $T_e$ to minimize the differences between the experimental spectral shape and calculated ones.

The least square value normalized by using the minimum least square value of each data is shown in Fig.\,\ref{fig:spectrum_result} (f,g,h,i). 
Fig.\,\ref{fig:spectrum_result} (f), (h), and (i) show the normalized least square values before the maximum compression time which correspond to Case A, C, and D in Table \ref{table: shot_summary}. 
In those cases, we can find the minimum least square value around 1.0 $\rm g/cm^3$. 
The spectra calculated with $T_e$ = 2.1, 1.9, and 1.5 keV, $\rho$ = 1.0 $\rm g/cm^3$, $d$ = 100 $\mu$m well reproduce the shape of the spectrum, as shown in Fig.\,\ref{fig:spectrum_result} (a), (c), and (d), respectively.
Assuming a smaller thickness and over the initial solid density, the temperature is found to be around 2 keV for Case B in Table \ref{table: shot_summary}, as shown in Fig.\,\ref{fig:spectrum_result}(g). 
The spectra calculated with $T_e$ = 2.0 keV, $\rho$ = 12 $\rm g/cm^3$, $d$ = 30 $\mu$m well reproduce the shape of the spectrum, see Fig.\,\ref{fig:spectrum_result}(b).
In case E which is minimum LFEX energy shot, the signal was too weak, therefore, it was impossible to evaluate the temperature by fitting. 

To identify the heating region, the heated core was imaged by Fresnel Phase Zone Plate(FPZP).
A target doped with titanium was heated with the LFEX laser and the $\mathrm{Ti-}K_\alpha$ and the $\mathrm{Ti-}He_\alpha$ emitted from the core were respectively imaged. 
When the RE beam passing through the cold plasma, $\mathrm{Ti-}K_\alpha$ photons are emitted. 
After the plasma is heated up, titanium is rapidly ionized and then emits $\mathrm{Ti-}He_\alpha$.
$\mathrm{Ti-}He_\alpha$ emissions indicating the heating region were strongly observed along the laser axis, so that the experimental result confirms that the heating region exists inside the core. Details of the visualization of heated plasma by FPZP including the observed images are presented in Methods section.

Our measurement showed that the core along the laser axis has 11.3 $\rm g/cm^3$ and part of the core has 2.0 keV at around maximum compression time which correspond to 2.2 PPa of UHED state. 
In our previous work, maximum laser-to-core coupling efficiency by the drag heating was 7.7 $\pm$ 1.2 $\%$ which corresponds to Case E in Table \ref{table: shot_summary}.
It means that 48 $\pm$ 7 J energy can be deposited in the core by the drag heating. 
Even at the maximum compression timing, the areal mass densities ($\rho L$) of the core along the RE beam path length ($L$) were $\rho L= 0.16 {\rm g/cm}^{2}$. Therefore, the energy deposition by the drag heating occurs over the entire core. 
The core was heated up to $T_e = 80 \pm 10$ eV only by the drag heating.
Our spectral analyses show that heated region is more localized and the temperature exceeds keV. It suggests that the other heating mechanism contributes predominantly to explain the spectral analyses.

We performed 2D-PIC simulations (PICLS \cite{Sentoku2008}) under an external magnetic field using the density distribution obtained in the experiment.
The heating laser is similar to the LFEX laser, which has wavelength, pulse shape, pulse duration, and peak intensity of the heating beams of 1.053 $\mu$m, Gaussian, 1.2 ps (FWHM), and $1.6 \times 10^{19}$ $\rm W/cm^{2}$.

\begin{figure}[ht]
 \centering 
 \includegraphics[width=80mm]{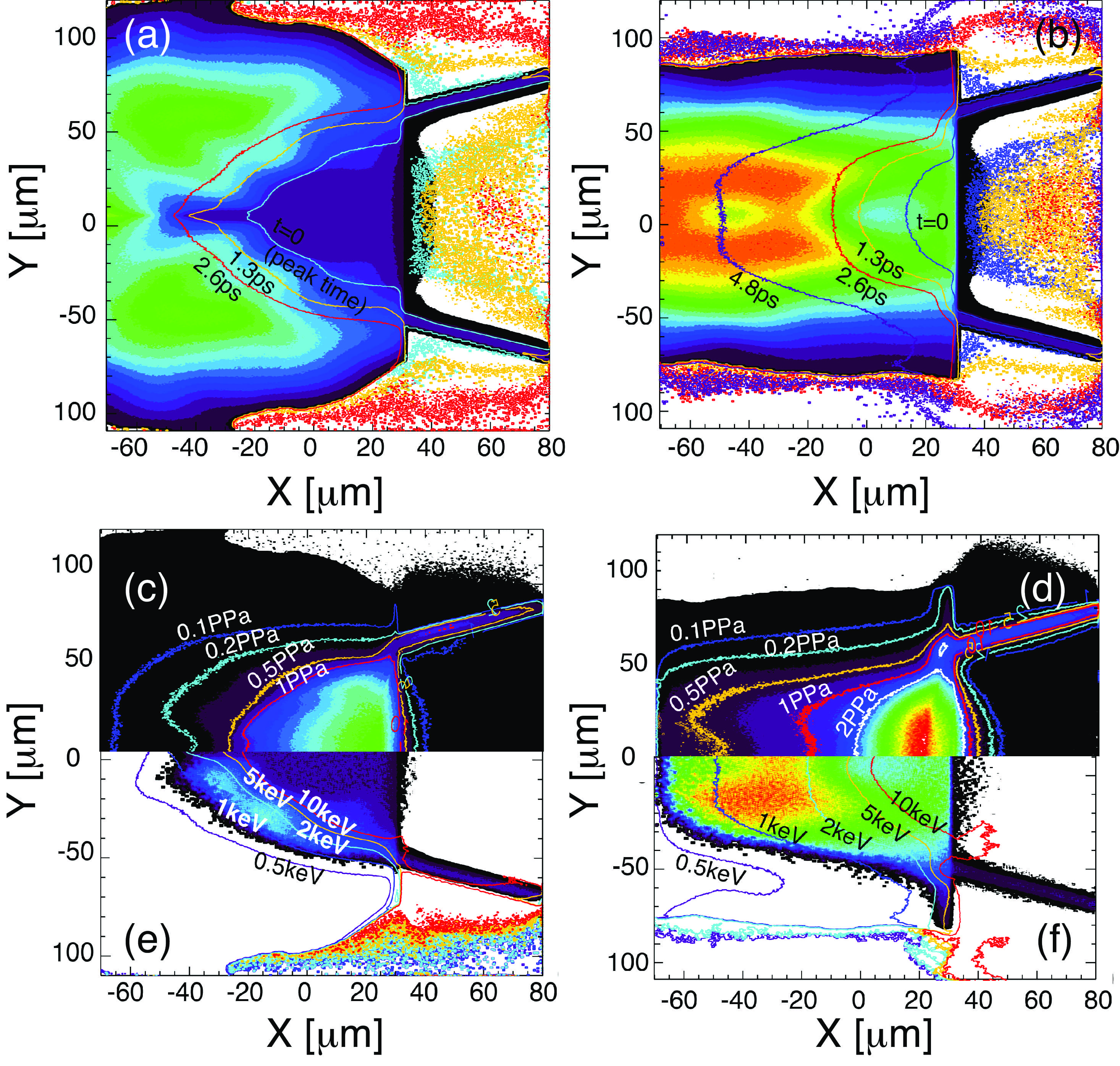}
\caption{(a), (b) The propagation of heat wave indicated by 1\,keV contour lines on the density profile. (c), (d) Pressure distributions with the contour lines (PPa). (e), (f) Electron temperature distribution on the density distribution of the doped copper with charge state $Z\ge 27$, which indicates where the $He_\alpha$ emissions come from. 
(a),(c),(e) are for the case before the maximum compression, and (c) and (e) are plotted at $t=2.6$\,ps.
(b),(d),(f) are for the case of the maximum compression, and (d) and (f) are plotted at $t=4.8$\,ps. 
The time is referred from the heating laser peak time.
} \label{fig:PIC_result}
\end{figure}

Fig.\,\ref{fig:PIC_result} (a) and (b) show the progress of heating indicated by countour lines of 1 keV from the heating laser peak time ($t=0$) for the cases with densities at $t_{\rm exp}$ = 0.38\,ns and 0.72\,ns in Fig.\,\ref{fig:density_map}.
The heating laser is irradiated from the right side through the cone and it heats the plasma near the core directly owing to the high contrast laser light.
The electron temperature evolves temporally via the thermal heat transport to the core by the diffusive heating. 
In the case of the maximum compression, Fig.\,\ref{fig:PIC_result} (b) and (f), the heat wave propagates even after the heating laser irradiation, and then the core region ($X<0$) was heated over 1\,keV electron temperature at $t=4.8$\,ps.

Fig.\,\ref{fig:PIC_result} (c) and (d) show the two dimensional pressure distribution at $t=2.6$\,ps and 4.8\,ps after heating laser peak time for cases (a) and (b). We see that the high pressure region propagates faster in (c) than that in (d) since the the plasma density before the maximum compression is lower.
The pressure of the core region ($X<0$) in (d) starts from 2\,PPa at the front edge to 0.5\,PPa at the other side ($X \simeq -50\,\mu$m) at the maximum compression.
Fig.\,\ref{fig:PIC_result} (e) and (f) show the bulk electron temperature on the density distribution of the doped copper having the charge states $Z\ge 27$. 
The doped copper densities with $Z\ge 27$ indicate that where $\mathrm{Cu-}He_\alpha$ photons are coming from. 
We see that the doped copper ions inside the core region in (f) get $Z\ge 27$, namely, the large amount of $He_\alpha$ emissions are expected from the core.
The core region at the maximum compression is heated to 1-2\,keV, which is consistent with the experimental observation, as seen in Fig.\,\ref{fig:spectrum_result} (g).
PIC simulations reveal that the diffusive heating is the heating process which can locally heat up the front region to the core region over peta-pascal pressure. 

Since the current core density is still low $\sim 10$\,g/cm$^3$, still two orders of magnitude less than from the ignition core, the RE drag heating does not play a significant role. 
While we find that the diffusive heating driven by the direct heating at the interaction surface by the heating laser is a key for creation of keV UHED plasma in the time scale of picoseconds.
Here, we revisit a simple one-dimensional model of diffusive heating.
In the diffusive process, a heat flux is proportional to the temperature gradient as
\begin{equation}
  \label{fourier}
  q=-\kappa \nabla T_e \approx \kappa \frac{T_b}{L},
\end{equation}
where $T_b$ is the bulk electron temperature near the solid surface, $L$ is the scale length of the temperature gradient. 
The heat conductivity coefficient $\kappa$ is assumed as $\kappa = n_{\rm e} v_{\rm th} l_{\rm mfp} = \kappa_{\rm SH} \times 3\pi/128 $ \cite{Asahina2019}, where $n_{\rm e}$ is the bulk electron density, $v_{\rm th}$ is the thermal velocity of $T_b$, $l_{\rm mfp}$ is the mean free path for
 $T_b$ and $n_{\rm e}$. $\kappa_{\rm SH}$ is the thermal conductivity in Spitzer-H\"{a}rm regime.
 We here assume that a part of laser energy flux is converted to the heat flux by the surface direct heating;
\begin{equation}
  \label{flux_conservation}
  \eta I = n_{\rm e} T_b v_{\rm heat},
\end{equation}
where $v_{\rm heat}$ is the propagation speed of the thermal conduction, and $\eta$ is the convergence ratio from the laser energy flux (intensity) $I$.
Hereafter we normalize each variable and express the normalized one with bar, i.e., velocity, density, time, temperature are normalized by the light speed $c$, critical density of laser light $n_{\rm c}$, laser period $\tau_{\rm L}$, electron rest mass energy $m_{\rm e}c^2$, respectively.
Using Eqs.\eqref{fourier} and \eqref{flux_conservation} and assuming that the scale length of diffusion increases as $\bar{L}\approx \bar{v}_{\rm heat} \bar{t}$, the bulk electron temperature near solid surface and the propagation speed of heat wave are derived as
\begin{equation}
  \bar{T}_{b} = \left( \bar{\Gamma}\ln \Lambda \left(\frac{\eta\,a_{\rm 0}^2}{2} \right)^2 \frac{\bar{t}}{\bar{n_{\rm e}}} \right)^\frac{2}{9},
\end{equation}
\begin{equation}
  \label{v_heat}
  \bar{v}_{\rm heat} = \left( \frac{1}{\bar{\Gamma}\ln \Lambda} \left(\frac{\eta\,a_{\rm 0}^2}{2} \right)^\frac{5}{2} \frac{1}{\bar{t}\,\bar{n_{\rm e}}^\frac{7}{2}} \right)^\frac{2}{9},
\end{equation}
where $\ln \Lambda$ is the Coulomb's logarithm, $\bar{\Gamma}=\Gamma n_{\rm c}\tau_{\rm L}\left(m_{\rm e}c^2\right)^{-3/2} = 2.65 \times 10^{-8}$. Here, $\Gamma=(4/3)(2\pi)^{1/2}e^4/m_e^{1/2}$ is a constant of the Coulomb's collision frequency.
Both $v_{\rm heat}$ and $T_b$ depend on the plasma density $n_e$, conversion rate $\eta$, and heating time $\bar{t}$, $v_{\rm heat}=13\,\mu$m/ps and $T_b=18$\,keV in 5\,g/cm$^{3}$ CH plasma with $\eta=0.3$, $\ln \Lambda=5$ at $t=2$\,ps. Note here that the time dependence is weak as $ \bar{T}_b \propto t^{2/9}$ and $\bar{v}_{\rm heat} \propto t^{-2/9}$. Those values are consistent with the simulations, as seen in Fig.\,\ref{fig:PIC_result} (b,d,f). 
From Eq.\eqref{v_heat}, the scale length of heated region is $\bar{L}\approx \bar{v}_{\rm heat} \bar{t}$ proportional to $(t / n_e)^{7/9}$.
This suggests that heating lasers with longer duration can heat denser core as a self-similar way.
The details of the theoretical model of the diffusive heating and comparison with PIC simulations will be discussed in another paper.


In summary, we have achieved experimentally 2.2 Peta-Pascal of UHED state with 4.6 kJ (1.5kJ is for compression, 1.7kJ is for generating a magnetic field, and 1.4kJ is for heating laser) of the total laser energy that is one order of magnitude lower than the conventional central implosion.
The generation of such the UHED state cannot be explained with the drag heating mechanism only, since the current core density is not sufficiently high to stop all of MeV electrons. 
Particle-in-cell simulations with the experimental conditions reveal that the diffusive heating mechanism plays an essential role to heat the core plasma over keV range on top of the drag heating and resistive heating.
In the ignition scale experiment, the REB drag heating is expected to play a significant role. 
As an example, we here assume that the core is about 10 times denser than the current experiment, i.e., $100\, {\rm g/cm}^3$, 
with the similar diameter $50\, {\rm \mu m}$, the heating laser has the same intensity and 10 times longer duration, thus 10 times higher energy than the current experiment, 
and the coupling efficiency from laser to the core electrons could be much larger than the current experiment $7.7\%$ for $\sim 10\, {\rm g/cm}^3$ core.
The core temperature is then estimated as $\sim 5\,{\rm keV}$ via the REB drag heating. 
While, the diffusive heating, which is proportional to the ratio of the heating time and the core density, will keep contributing to the heating. 
Our experimental results clarified that the magnetized fast isochoric heating is an efficient way to create the Peta-Pascal UHED state that is an interesting and unique testbed for various scientific researches, e.g., inertial confinement fusion, intense X-ray, charged particles, and neutron sources, and laboratory astrophysics.

\begin{acknowledgments}
The authors thank the technical support staff of ILE and the Cyber Media Center at Osaka University for assistance with the laser operation, target fabrication, plasma diagnostics, and computer simulations. 
This work was supported by the Collaboration Research Program between the National Institute for Fusion Science and the Institute of Laser Engineering at Osaka University, and by the Japanese Ministry of Education, Science, Sports, and Culture through Grants-in-Aid, KAKENHI (Grants No. 24684044, 25630419, 15K17798, 15K21767, 15KK0163, 16K13918, 16H02245, and 17K05728), Bilateral Program for Supporting International Joint Research by JSPS, and Grants-in-Aid for Fellows by Japan Society for The Promotion of Science (Grant No. 14J06592, 15J00850, 15J00902, 15J02622, 17J07212, 18J01627, 18J11119, and 18J11354). 
This research used computational resources of the HPCI system provided by (Information Technology Center, Nagoya University) through the HPCI System Research Project (Project ID: hp180093).
The study also benefited from diagnostic support funded by the French state through research projects TERRE ANR-2011-BS04-014 (French National Agency for Research (ANR) and Competitiveness Cluster ”Route des Lasers”) and ARIEL (Regional Council of Aquitaine). M. B.-G. and J. J. S. acknowledge the financial support received from the French state and managed by ANR in the framework of the ”Investments For the Future” program at IdEx Bordeaux - LAPHIA (ANR-10-IDEX-03-02), from COST Action MP1208 ”Developing the Physics and the Scientific Community for Inertial Fusion” and from the EUROfusion Consortium and has received funding from the Euratom research and training programs 2014-2018 under grant agreement No 633053. The views and opinions expressed herein do not necessarily reflect those of the European Commission.
\end{acknowledgments}

\section{Author Contributions}
S. F. is the principal investigator who proposed and organized this study. 
K. M., S. S., and S. L. are equally contribute to the experiment and analysis. 
N. H.,  N. I. and Y.S built a model of diffusive heating.
The experiment was carried out with the help of H. M., K. F. F. L., A. Y., Y. O., Y. A., and S. K. under the supervisions by H. S., A. M., A. Y., M. N., T. O., H. N., H. A., R. K., Y. A., and S. F.. 
K. M. and S. S. estimate the electron temperature of the core from $\mathrm{Cu}$ spectrum.
S. L., and H. S. analyzed the backlight images. 
K. M. measure the $\mathrm{Ti-}K_\alpha$ and $\mathrm{Ti-}He_\alpha$ distribution by using FPZP.
Y. I., K. Y. and T.N. developed targets used in the experiment. 
Y. N., S. T., N. M., and J. K. contributed to improve intensity contrast of the LFEX laser pulses. 
T. J., M. H., A. S., T. S., H. S., K. M., H. N., and Y. S. performed computer simulation and theoretical analysis. 
M. B-G., and J. J. S. contributed to design of the integrated MFI experiment based on their experimental results obtained in the LULI2000 facility. All authors contributed to the discussion of the results. 

\section{Author Information}
The authors declare no competing interests.
Correspondence and requests for materials should be addressed to K. M. (matsuo-k@ile.osaka-u.ac.jp) or S. F. (sfujioka@ile.osaka-u.ac.jp).

\bibliographystyle{naturemag}
\bibliography{library}

\section*{Methods}

\subsection{Quasi-monochromatic imaging of diffusively heated region}
The idea to use a diffraction for recording hard X-ray images has been published in 1963 \cite{Denisiuk1963}.
The application of this idea known as Fresnel Phase Zone Plate(FPZP) designed for focusing and imaging hard X-ray \cite{Kirz1973, Do2018}.
In our case, it was used to image $\mathrm{Ti-}K_\alpha$ and $\mathrm{Ti-}He_\alpha$ emissions from the isochorically heated region. 

Our FPZP is consists of a multiple circular transmission grating made of tantalum.
A phase difference occurs between X-rays passed through the transparent zones and passed through the grating zones. To chose an appropriate thickness, these zones produce a phase shift of $\pi$, therefore contributing to constructive interferences at the lens focus.

X-ray is focused at the point of the focal length $f= a^2/\lambda$ with the X-ray wavelength $\lambda$ and the FPZP innermost ring radius $a$. The innermost ring radius $a$ of our FPZP is 6 $\mu$m, therefore the focal distance for $\mathrm{Ti-}K_\alpha$ is 131mm, and for $\mathrm{Ti-}He_\alpha$ is 136mm, respectively. In principle imaged by FPZP can be calculated according to the following expression,
\begin{equation}
    \frac{1}{f}=\frac{1}{r}+\frac{1}{R}
\end{equation}
where $r$ is the distance between the X-ray source and the FPZP, $R$ is the distance between the FPZP and the X-ray detector. The focal length is a function of the wavelength, therefore the wavelength differs shifts the focal position of the image. FPZP itself has wavelength selectivity. The spatial resolution, and spectral bandwidth were 11 $\mu$m (FWHM), and 120 eV (FWHM), respectively.

We measured the two-dimensional heating distribution of the fast heated high energy density plasma.
A heavy water target containing titanium as a tracer is used for measuring the two-dimensional heating distribution. The 250 $\mu$m diameter, 10 $\mu$m thickness plastic shell is filled with D$_2$O liquid contained 2.3$\%$ Ti atoms in weight through filltube \cite{Iwasa2018}. A open-tip gold cone is attached to this heavy water target. The heavy water target was compressed by GEKKO beams and heated by LFEX beams.

\begin{figure}[h]
 \centering 
 \includegraphics[width=90mm]{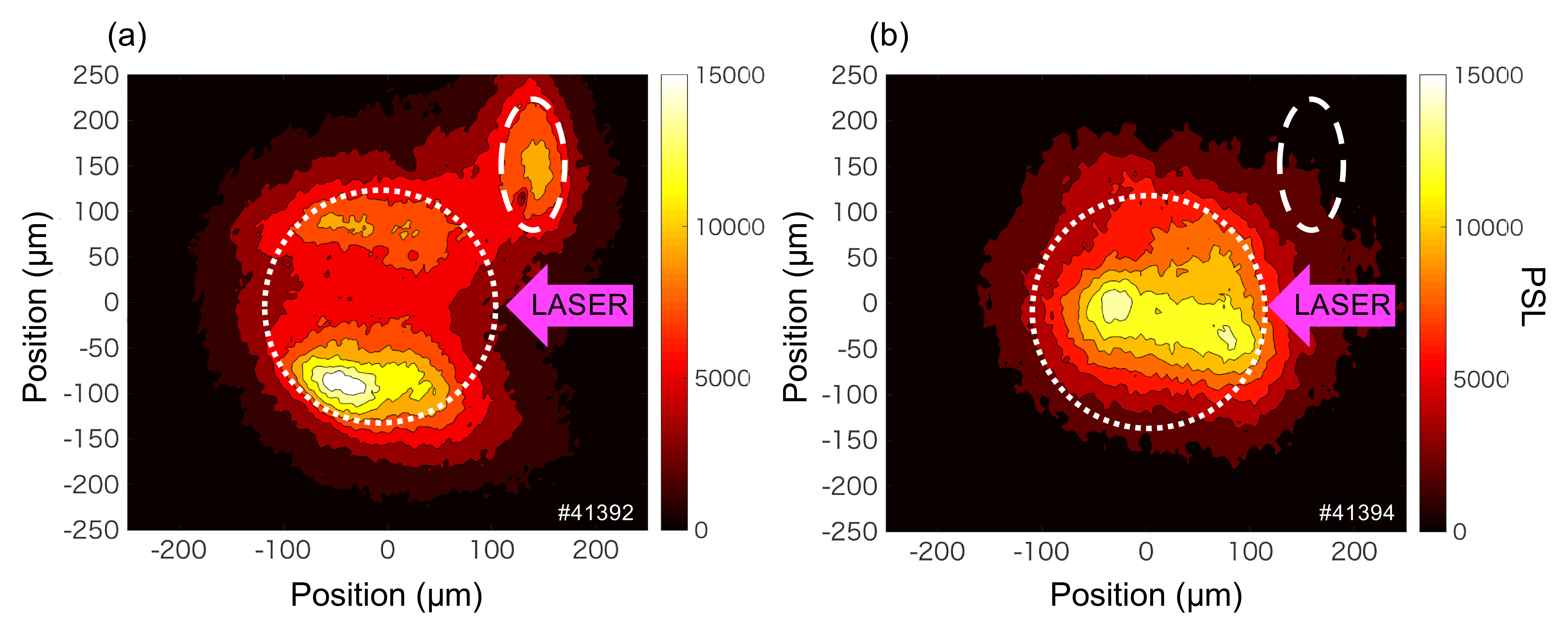}
 \caption{(a)The $\mathrm{Ti-}K_\alpha$ emission and (b)$\mathrm{Ti-}He_\alpha$ emission from the core. A dotted line show the initial diameter of the target and a dashed line shows the position of the filltube. The $\mathrm{Ti-}K_\alpha$ emission is observed edge of the target and inside of filltube. The $\mathrm{Ti-}He_\alpha$ emission where is the heating region was strongly observed along the laser axis. } \label{fig:FPZP_result}
\end{figure}

The $\mathrm{Ti-}K_\alpha$ and the $\mathrm{Ti-}He_\alpha$ emitted from the core were respectively imaged by FPZP to identify the heating region, as shown in Fig.\,\ref{fig:FPZP_result}.
When the REB passing through the cold plasma, $\mathrm{Ti-}K_\alpha$ is emitted. The $\mathrm{Ti-}K_\alpha$ emission is observed edge of the target and inside of filltube, as shown in Fig.\,\ref{fig:FPZP_result} (a). 
Titanium is rapidly ionized after heated up, and can emit He-like X-ray such as $\mathrm{Ti-}He_\alpha$.The $\mathrm{Ti-}He_\alpha$ emission where is the heating region was strongly observed along the laser axis, as shown in Fig.\,\ref{fig:FPZP_result} (b). The difference of emission region shows that the core is locally heated along the laser axis.

\subsection{Numerical simulations}
To study the LFEX heating of imploded plasmas, we had performed Particle-in-Cell (PIC) simulations. 
Simulation results in 2D geometry shown in Fig.\,3 are obtained using a fully relativistic PIC code PICLS, which features binary collisions among charged particles and dynamic ionization \cite{Sentoku2008}. 
The size of simulation box is 150\,$\mu$m in the laser propagation direction and 228\,$\mu$m in the transverse direction with the mesh size of $1/35\,\mu$m. 
To avoid the numerical heating, the fourth-order interpolation scheme is applied. 
Absorbing boundary conditions are used for particles in the transverse direction (i.e., no electron reflux is imposed to represent the actual large transverse size of the target). 
The laser pulse profile and the wavelength are similar as those of the LFEX laser.
The plasma density profiles are set by the measured plasma density shown in Fig.\,1 (a) for a case before the maximum compression and (b) for the maximum density compression.
The target consists a neutral plasma of electrons and C$^{3+}$, containing 5\% Cu$^{3+}$ ions in weight. 
The number of particles per cell is 32 (29 for electrons and 3 for ions) and the total number of particles used in the simulation is about 1.4 billion. 
A uniform external magnetic field with 1\,kT along X-axis is applied as in the experiment. 

\end{document}